\def\tr{{\rm tr}\,}

\def\sgn{{\rm sgn\,}}
\def\b{\bibitem}
\def\be{\begin{equation}}
\def\ee{\end{equation}}
\def\bea{\begin{eqnarray}}
\def\eea{\end{eqnarray}}
\def\bml{\begin{mathletters}}
\def\eml{\end{mathletters}}
\documentstyle[aps,prb,eqsecnum,psfig,epsf,floats]{revtex}
\draft
\begin{document}
\def\SNG{{\em Physical Review Style and Notation Guide}}
\def\LUG {{\em \LaTeX{} User's Guide \& Reference Manual}}
\def\btt#1{{\tt$\backslash$\string#1}}%
\def\REVTeX{REV\TeX}
\def\AmS{{\protect\the\textfont2
        A\kern-.1667em\lower.5ex\hbox{M}\kern-.125emS}}
\def\AmSLaTeX{\AmS-\LaTeX}
\def\BibTeX{\rm B{\sc ib}\TeX}
\twocolumn[\hsize\textwidth\columnwidth\hsize\csname@twocolumnfalse%
\endcsname
\title{Electrons in an annealed environment:\\
       A special case of the interacting electron problem
}
\author{D.Belitz}
\address{Department of Physics and Materials Science Institute\\
         University of Oregon,
         Eugene, OR 97403}
\author{T.R.Kirkpatrick}
\address{Institute for Physical Science and Technology, and Department of
         Physics\\
         University of Maryland,
         College Park, MD 20742}
\date{\today}
\maketitle

\begin{abstract}
The problem of noninteracting electrons in the presence of annealed
magnetic disorder, in addition to nonmagnetic quenched disorder, 
is considered. 
It is shown that the proper physical interpretation of this model is
one of electrons interacting via a potential that is long-ranged in
time, and that its technical analysis by means of renormalization
group techniques must also be done in analogy to the interacting
problem. As a result, and contrary to previous claims, the model
does not simply describe a metal-insulator transition in
$d=2+\epsilon$ ($\epsilon\ll 1$) dimensions. Rather, it describes
a transition to a ferromagnetic state that, as a function of the
disorder, precedes the metal-insulator transition close to $d=2$.
In $d=3$, a transition from a paramagnetic metal to a paramagnetic
insulator is possible.
\end{abstract}
]
\section{Introduction}
\label{sec:I}

Local magnetic moments are known to play an important role in the behavior
of disordered electronic systems, but the precise nature of that role
remains incompletely understood.\cite{us_R} 
One way to think about such local moments
is that, in a disordered environment, the exchange interaction between
the electrons may be locally enhanced to the point where the electron
spins order magnetically in a finite region in space.\cite{LM_footnote} 
The resulting magnetized regions
are often referred to as local moments, or droplets, or rare regions.
Since they are self-generated by the electron system, they are in
thermodynamic equilibrium with the other electronic degrees of freedom.
It is therefore intuitively plausible that such local moments can be
modeled as annealed magnetic disorder, in addition to the underlying
quenched disorder that produces them. In Ref.\ \onlinecite{us_letter}
an explicit derivation has been given that corroborates this argument.
There is experimental evidence for such local moments to influence the
transport properties of the electron system in important ways, and in
particular they are suspected to influence the critical behavior near
the metal-insulator transition (MIT) that is observed in disordered electron
systems.\cite{us_R} However, theoretically understanding the coupling between 
local moments and transport properties has proven to be very hard. Studying
and understanding the annealed disorder model mentioned above is expected
to shed light on this important problem.

Reference \onlinecite{us_letter} provided such an analysis, and concluded
that the annealed disorder leads to a new and very interesting type of
MIT. The most exciting feature was that the
transition was driven by the vanishing of the thermodynamic density
susceptibility $\partial n/\partial\mu$, and thus resembled a Mott
transition more than an Anderson transition.\cite{Mott} This was even more
surprising as the Coulomb interaction between the electrons, which
is what usually causes a Mott transition, had not been explicitly taken into
account in the model.

Subsequently, Ref.\ \onlinecite{us_local_OP} developed a general classification
of quantum phase transitions with respect to, (1) whether one can describe the
transition by means of a local order parameter, and (2) whether the order
parameter susceptibility in the disordered phase is an analytic function
of the wavenumber. The second criterion has an important bearing on which
observables can become critical at a MIT: Criticality
in $d>2$ ($d=2$ is the lower critical dimension for all known MITs
of disordered interacting electrons) implies a logarithmic
dependence on the renormalization group (RG) length rescaling factor, and hence
on the wavenumber, in perturbation theory in $d=2$. This in turn implies
a (weaker) nonanalytic wavenumber dependence in $d>2$ away from 
criticality.\cite{us_chi_s,NLsM_footnote} Although
the considerations in Ref.\ \onlinecite{us_local_OP} do not provide a 
mathematically rigorous proof, they strongly suggest that 
$\partial n/\partial\mu$ cannot be critical at a MIT for a large class of 
models, which includes the model studied in Ref.\ \onlinecite{us_letter}.

In the current paper we provide a thorough re-analysis of the model derived
and motivated in Ref.\ \onlinecite{us_letter}, and resolve this contradiction.
We show that the RG analysis of the model performed in
Ref.\ \onlinecite{us_letter} had an incorrect structure and led to unreliable
results. A proper analysis of the model's renormalizability, and the
resulting RG flow equations, show that 
$\partial n/\partial\mu$ is not singularly renormalized and hence not
critical, in agreement with Ref.\ \onlinecite{us_local_OP}. 
In addition, it reveals that within a controlled $\epsilon$-expansion
about $d=2$, the model does not simply describe a metal-insulator 
transition. Rather, it displays a variant of the phase transition sequence 
that is known to occur in a related model with both quenched disorder and 
electron-electron interactions (but no annealed disorder).\cite{us_IFS} 
That is, as the disorder increases, there is first a transition to a 
ferromagnetic metallic state, and then, with further increasing disorder, 
a transition to a ferromagnetic insulator state. For $d=3$ a transition
directly from a paramagnetic metal to a paramagnetic insulator is possible.

This paper is organized as follows. In the next section, we give intuitive
physical arguments that explain our model and our procedure to analyze it,
and we summarize our results. In Sec.\ \ref{sec:III} we formally define the
model and write it in a way that facilitates a renormalization group analysis.
Sec.\ \ref{sec:IV} performs the renormalization to one-loop order, and
Sec.\ \ref{sec:V} analyzes the results. Some technical issues regarding the
model's renormalization properties are relegated to Appendices A and B, the
flow equations for the interacting and annealed disorder models are compared
in Appendix C, and a perturbative analysis of the free energy is given in
Appendix D.

\section{Physical arguments}
\label{sec:II}

Since some of our detailed arguments are quite technical, we start 
by giving some intuitive physical arguments to explain both
our general strategy and our results.

\subsection{Annealed disorder as a model for local moments}
\label{subsec:II.A}

We start by recalling the argument for why annealed disorder models local
moments.\cite{us_letter} Any field theoretical treatment of a statistical
mechanics problem starts with a functional integral representation of the
partition function,\cite{ItzyksonDrouffe}
\bml
\label{eqs:2.1}
\be
Z = \int D[\phi]\ e^{-S[\phi]}\quad.
\label{eq:2.1a}
\ee
The form of the action $S$ defines the model under consideration, and
the mathematical nature of the field $\phi$ depends on whether the
system is classical or quantum mechanical, consists of fermions or bosons, 
and whether the model is a
microscopic one in terms of fundamental fields, or of an effective
nature. The usual procedure is to identify a saddle point of $S$
that approximately contains the physics one is interested in, to
expand about this saddle point, and to employ perturbation theory
and the renormalization group. In a system with quenched disorder
there will be, apart from homogeneous saddle-point solutions, solutions
where the field $\phi$, or some components of it, are nonzero only
in certain regions in space. Such inhomogeneous saddle points have
been proposed as a description of rare regions in classical magnets 
by Dotsenko et al.\cite{LM_footnote} This concept was 
generalized to quantum magnets,\cite{us_rr_magnets} and to the effective
field theories used to describe MITs in quenched
disordered electron systems.\cite{us_letter} In a large system there will be 
many rare regions that interact only very weakly, 
and thus exponentially many almost degenerate saddle
points, since the orientation of the field on the rare regions is arbitrary.
These saddle points are expected to be separated by large energy barriers,
and thus to not be perturbatively accessible from one another. Within
perturbation theory, and denoting the $n$-th saddle-point field configuration
by $\Phi^{(n)}$ and the fluctuations by $\varphi$, one can therefore
write the partition function
\be
Z \approx \sum_{n} D[\varphi]\ e^{-S[\Phi^{(n)} + \varphi]}\quad.
\label{eq:2.1b}
\ee
In the thermodynamic limit, the discrete set of saddle points becomes a
saddle-point manifold that needs to be integrated over. The saddle-point
field configurations $\Phi$ thus become degrees of freedom that are governed by
some probability distribution $P[\Phi]$, are integrated
over at the level of the partition function, and couple to the field $\varphi$
by means of some coupling $S_{\rm c}$ that is determined by the action $S$,
\be
Z \approx \int D[\Phi]\ P[\Phi] \int D[\varphi]\ 
          e^{-S[\varphi] + S_{\rm c}[\Phi,\varphi]}\quad.
\label{eq:2.1c}
\ee
\eml%
They therefore act like annealed disorder.
Note that in giving Eq.\ (\ref{eq:2.1c}) we implicitly assume that the
(annealed) disorder adjusts and comes to equilibrium with the
fluctuations $\varphi$. If the disorder were fixed on the time scale of
the $\varphi$-fluctuations, then it would be quenched disorder. In the
latter case, for the average over saddle-points to be meaningful, $\ln Z$ 
rather than $Z$ should be averaged over the $\Phi$-fields.\cite{Grinstein}

In our case we are interested in rare regions that carry a magnetic moment.
According to the arguments recalled above, they can be modeled by annealed
magnetic disorder in addition to the quenched disorder that allows for the
inhomogeneous saddle-point solutions. In the simplest possible model the
annealed disorder has a Gaussian distribution, and is static. The latter
means that the coupling constant, or the annealed magnetic disorder
strength, will be proportional to the temperature.\cite{us_letter} 
This is just the
Boltzmann weight assigned to these classical degrees of freedom
that are in equilibrium with the electrons. We emphasize that this model,
and its derivation in Ref.\ \onlinecite{us_letter}, is unaffected by
our considerations concerning its analysis and interpretation, which differ
from the one given in that reference.

\subsection{Annealed disorder as an effective interaction}
\label{subsec:II.B}

The physical effects of annealed disorder are fundamentally different from 
those of quenched, or frozen-in, disorder.\cite{Grinstein} 
The former gets integrated over at the level of the partition
function, cf. Eq.\ (\ref{eq:2.1c}), the latter, at the level of the
free energy. Consequently, integrating out annealed disorder generates
a physical effective interaction between the degrees of freedom that couple
to it, the electron spin density in our case, 
which can be understood as resulting from an
exchange of annealed disorder fluctuations between the electrons. The
effects of quenched disorder, on the other hand, are more subtle and
fundamentally different from those of interactions. 

It is therefore plausible that a system of noninteracting electrons in the 
presence of both quenched and annealed disorder will behave in many respects 
like one with quenched disorder only and an additional electron-electron 
interaction. As the only difference one would expect that, if the
annealed disorder is modeled as static, the resulting effective interaction
will be infinitely long-ranged in time, a feature that one would not
expect to have qualitative effects. This expectation is in contradition
with the results of Ref.\ \onlinecite{us_letter}, which found behavior that
was drastically different from that of electrons interacting via an
instantaneous interaction. In particular, this reference predicted a
MIT of Mott type, where the thermodynamic
susceptibility $\partial n/\partial\mu$ vanishes. This is in contradiction to
both explicit calculations for quenched disordered, interacting electron
systems, which find that $\partial n/\partial\mu$ is not singularly
renormalized,\cite{F,us_R} and very general considerations in 
Ref.\ \onlinecite{us_local_OP}. 

The analysis that will be presented below removes this contradiction,
and illustrates the technical issues behind the above intuitive
physical considerations. We will show that the technical treatment of the 
annealed disorder in analogy to that of quenched disorder in Ref.
\onlinecite{us_letter} was not only in disagreement with the above physical
arguments, but led to an unnatural structure of the theory. This in turn
led to incorrect assumptions about the behavior under renormalization,
and ultimately to physically incorrect results. A treatment of the
annealed disorder in analogy to an interaction, on the other hand,
does not run into these problems and yields results that are in agreement
with all known constraints.

\section{The model and its renormalizability}
\label{sec:III}

In this section we consider the same effective field theory as in
Ref.\ \onlinecite{us_letter}.

\subsection{Effective field theory}
\label{subsec:III.A}

Our starting point, as in Ref.\ \onlinecite{us_letter}, is
Wegner's nonlinear sigma-model
(NL$\sigma$M)\cite{Wegner} for noninteracting electrons with nonmagnetic
quenched disorder. The action reads
\be
{\cal A}_{{\rm NL}\sigma{\rm M}} = \frac{-1}{2G}\hskip -1pt\int\hskip -1pt 
   d{\bf x}\,\tr \left[\nabla
   Q({\bf x})\right]^2 + 2H^{(1)}\hskip -3pt\int\hskip -2pt 
   d{\bf x}\,\tr \left[\Omega\,Q({\bf x}) \right].
\label{eq:3.1}
\ee
Here $Q({\bf x})$ is a matrix field that comprises two fermionic degrees of
freedom. Accordingly, $Q$ carries two
fermionic Matsubara frequency indices $n$ and
$m$, and two replica indices $\alpha$ and $\beta$ to deal with the quenched
disorder. The matrix elements $Q_{nm}^{\alpha\beta}$ are spin-quaternion
valued to allow for particle-hole and spin degrees of freedom. It is
convenient to expand them in a basis $\tau_r\otimes s_i$ ($r,i=0,1,3.3$)
where $\tau_0=s_0$ is the $2\times 2$ unit matrix, and
$\tau_{1,3.3} = - s_{1,3.3} = -i\sigma_{1,3.3}$, with $\sigma_j$ the Pauli
matrices,\cite{ELK}
\bml
\label{eqs:3.2}
\be
Q_{nm}^{\alpha\beta} = \sum_{r}\sum_{i} {^i_rQ}_{nm}^{\alpha\beta}\quad.
\label{eq:3.2a}
\ee
For simplicity, we will ignore the particle-particle or Cooper
channel, which amounts to dropping $\tau_1$ and $\tau_2$ from the
spin-quaternion basis.\cite{ELK,us_R} The $Q_{nm}^{\alpha\beta}$ are then
elements of ${\cal C}\times{\cal Q}$, with ${\cal C}$ and ${\cal Q}$ the
complex number field and the quaternion field, respectively. 
The ${^i_rQ}_{nm}^{\alpha\beta}$ obey the following symmetry properties
(for $r=0,3$),\cite{us_fermions}
\bea
{^0_rQ}_{nm}^{\alpha\beta} &=& (-)^r\ {^0_rQ}_{mn}^{\beta\alpha}\quad,
\label{eq:3.2b}\\
{^i_rQ}_{nm}^{\alpha\beta} &=& (-)^{r+1}\ {^i_rQ}_{mn}^{\beta\alpha}\quad,\quad
                             (i=1,3.3)\quad.
\label{eq:3.2c}
\eea
Alternatively,
we can write the spin indices explicitly, and consider matrix elements
$Q_{nm,ij}^{\alpha\beta}$ that are complex number valued. $Q$ is subject 
to the constraints
\be
Q^2({\bf x}) \equiv 1\quad,\quad \tr Q({\bf x}) \equiv 0\quad.
\label{eq:3.2d}
\ee
These constraints are conveniently implemented by parametrizing $Q$ in
terms of matrices $q$ whose matrix elements, $q_{nm}^{\alpha\beta}$,
are restricted to frequency labels $n>0$, $m<0$. In terms of the $q$,
$Q$ can be written in block matrix form 
\be
Q = \left(\matrix{\sqrt{1-qq^{\dagger}} - 1 & q \cr
                  q^{\dagger} & -\sqrt{1 - q^{\dagger}q} + 1\cr}\right)\quad.
\label{eq:3.2e}
\ee
\eml%
Here the block matrices, clockwise from the upper left, correspond to frequency 
labels $n,m>0$; $n>0$, $m<0$; $n,m<0$; and $n<0$, $m>0$, respectively.

$\Omega_{nm}^{\alpha\beta} = \delta_{nm}\delta_{\alpha\beta}\Omega_n\,
(\tau_0\otimes s_0)$ in Eq.\ (\ref{eq:3.1})
is a frequency matrix with $\Omega_n = 2\pi Tn$ a bosonic Matsubara frequency
and $T$ the temperature.
$G$ is a measure of the disorder that is proportional
to the bare resistivity, and the frequency coupling $H^{(1)}$ 
is proportional to
the bare density of states at the Fermi level. $\tr$ denotes a trace over
all discrete degrees of freedom that are not shown explicitly.

The properties of this model are well known.\cite{Wegner,ELK,us_R}
The bare action describes diffusive electrons, with $D=1/GH^{(1)}$ 
the diffusion coefficient.
Under renormalization, $D$ decreases with increasing
disorder until a MIT is reached at a critical disorder value. The critical
behavior is known in an $\epsilon$-expansion about the lower
critical dimension $d=2$. In the absence of the Cooper channel, the MIT
appears only at two-loop order at a critical disorder strength of
$O(\sqrt\epsilon)$. $H^{(1)}$, which determines the specific heat coefficient,
the spin susceptibility, and $\partial n/\partial\mu$, is uncritical,
which makes this MIT an Anderson transition.

Now we add magnetic annealed disorder to the model. The motivation for
this is the fact that annealed disorder models certain types of local
moments, see Secs.\ \ref{sec:I} and \ref{sec:II} above. A technical derivation
of this has been given in Ref.\ \onlinecite{us_letter}, and the main idea 
has been recapitulated in Sec.\ \ref{subsec:II.A}. 
Annealed disorder implies that
the $Q$ in the resulting terms all carry the same replica 
index;\cite{Grinstein} otherwise, the functional form of the resulting 
additional term in the action can be taken from Ref.\ \onlinecite{ELK},
which considered quenched magnetic disorder. From that reference, we have
\bml
\label{eqs:3.3}
\be
{\cal A}_{\rm ann}^{(1)} = 2TJ^{(1)} \int d{\bf x}\sum_{\alpha}\sum_{j=1}^{3}\tr
                     \left[(\tau_3\otimes s_j)\,Q^{\alpha\alpha}({\bf x})
                     \right]^2\quad.
\label{eq:3.3a}
\ee
The coupling constant $J^{(1)}$ is a measure of the strength of the magnetic
disorder. The temperature prefactor in Eq.\ (\ref{eq:3.3a}) is a consequence
of the static nature of the local moments considered within this model, as
has been explained in Ref.\ \onlinecite{us_letter} and Sec.\ \ref{sec:II}
above. Equation (\ref{eq:3.3a}) is the only annealed magnetic disorder term if
fluctuations of the matrix field $Q$ on all length scales are taken into
account in calculating the partition function. However, the NL$\sigma$M
is an effective theory for long-wavelength fluctuations, and it is therefore
convenient to project the annealed disorder term onto this regime as well.
It has been discussed in detail in Ref.\ \onlinecite{us_fermions} that this 
can be achieved by means of a phase space decomposition and a relabeling of
momenta. Applied to Eq.\ (\ref{eq:3.3a}), this procedure generates another
contribution to the action,
\be
{\cal A}_{\rm ann}^{(2)} = 2TJ^{(2)} \int d{\bf x}\sum_{\alpha}\sum_{j=1}^{3}
                     \left[\tr (\tau_3\otimes s_j)\,Q^{\alpha\alpha}({\bf x})
                     \right]^2\quad.
\label{eq:3.3b}
\ee
\eml%
The coupling constant $J^{(2)}$ is in general independent of $J^{(1)}$.
${\cal A}_{\rm ann}^{(1)}$ and ${\cal A}_{\rm ann}^{(2)}$ enter the action
additively with the understanding that only long-wavelength fluctuations
are integrated over in calculating the partition function.
Note that in the case of quenched magnetic disorder, a complete phase space 
decomposition leads to a term analogous to Eq.\ (\ref{eq:3.3b}), 
but it is zero in the replica limit because the replica sum is then part of 
the trace.

As we will see, under renormalization the annealed disorder terms generate
another contribution to the action that takes the form
\be
{\cal A}_{\Omega}^{(2)} = 2H^{(2)}\int d{\bf x}\ 
                \tr\left[\sgn\Omega\ Q({\bf x})\right]\quad,
\label{eq:3.4}
\ee
so we add this right away. For a discussion on why this term must be
present on physical grounds, see Section \ref{subsec:V.B}.
\be
{\cal A} = {\cal A}_{{\rm NL}\sigma{\rm M}} + {\cal A}_{\rm ann}^{(1)}
           + {\cal A}_{\rm ann}^{(2)} + {\cal A}_{\Omega}^{(2)}\quad,
\label{eq:3.5}
\ee
is the complete action for our model, and the partition function is
obtained as the functional integral
\be
Z = \int D[Q]\ \delta[Q^2-1]\ e^{{\cal A}[Q]}\quad.
\label{eq:3.6}
\ee

\subsection{Annealed disorder as a long-ranged interaction}
\label{subsec:III.B}

${\cal A} = {\cal A}_{{\rm NL}\sigma{\rm M}} + {\cal A}_{\rm ann}^{(1)}$
defines the model studied
in Ref.\ \onlinecite{us_letter}.
${\cal A}_{\rm ann}^{(2)}$ was neglected in that reference, but this
term will not be of crucial importance in what follows. Terms that
appear under renormalization and indicate the appearance of
${\cal A}_{\Omega}^{(2)}$ were interpreted differently in Ref.\ 
\onlinecite{us_letter}, and we will discuss this point in 
Sec.\ \ref{subsec:V.B} below. A related point is that we have written
${\cal A}_{\rm ann}^{(1)}$ in a form that is different
from the one in Ref.\ \onlinecite{us_letter}. The latter
representation was modeled after the way one would treat quenched
disorder, and it added and subtracted a term where all replica indices
of the $Q$ are not the same. As we will see, this formulation, which
is a matter of taste at this point, is rather unnatural at the stage
of a RG analysis, and this led to the incorrect RG treatment of the model
in Ref.\ \onlinecite{us_letter}. We therefore write the annealed disorder
term in a form that is strictly diagonal in the replica index. This
replica structure is common to both the annealed disorder term, and
any electron-electron interaction term, and one would therefore expect
the renormalization properties of the current model and one of interacting
electrons to have common features. To underscore this point, we rewrite
the annealed disorder part of the action by splitting it into spin-singlet
and spin-triplet contributions,
\bml
\label{eqs:3.7}
\be
{\cal A}_{\rm ann} \equiv {\cal A}_{\rm ann}^{(1)} + {\cal A}_{\rm ann}^{(2)}
\equiv {\cal A}_{\rm ann}^{(1,s)} + {\cal A}_{\rm ann}^{(1,t)}
       + {\cal A}_{\rm ann}^{(2,t)}\quad,
\label{eq:3.7a}
\ee
with
\bea
{\cal A}_{\rm ann}^{(1,s)}&=&\frac{-\pi T}{4}\,J^{(1,s)}\sum_{nm}\sum_{\alpha}
   \sum_{r=0,3} (-)^r 
\nonumber\\
&&\hskip 0pt \times\tr \left[(\tau_r\otimes s_0)\,Q_{nm}^{\alpha\alpha}
   ({\bf x})\right]
\tr\left[(\tau_r\otimes s_0)\,Q_{mn}^{\alpha\alpha}
   ({\bf x})\right]\quad,
\nonumber\\
\label{eq:3.7b}
\eea
\bea
{\cal A}_{\rm ann}^{(1,t)}&=&\frac{\pi T}{4}\,J^{(1,t)}\sum_{nm}\sum_{\alpha}
   \sum_{r=0,3} (-)^r \sum_{i=1}^{3}
\nonumber\\
&&\hskip 0pt \times\tr \left[(\tau_r\otimes s_i)\,Q_{nm}^{\alpha\alpha}
   ({\bf x})\right]
\tr\left[(\tau_r\otimes s_i)\,Q_{mn}^{\alpha\alpha}
   ({\bf x})\right]\quad,
\nonumber\\
\label{eq:3.7c}
\eea
\bea
{\cal A}_{\rm ann}^{(2,t)}&=&\frac{-\pi T}{4}\,J^{(2,t)}\sum_{nm}\sum_{\alpha}
   \sum_{r=0,3} (-)^r \sum_{i=1}^{3}
\nonumber\\
&&\hskip 0pt \times\tr \left[(\tau_r\otimes s_i)\,Q_{nn}^{\alpha\alpha}
   ({\bf x})\right]
\tr\left[(\tau_r\otimes s_i)\,Q_{mm}^{\alpha\alpha}
   ({\bf x})\right]\quad,
\nonumber\\
\label{eq:3.7d}
\eea
where we have used Eqs.\ (\ref{eq:3.2a}) - (\ref{eq:3.2c}).
Here 
\be
J^{\rm (2,t)} = 8J^{(2)}/\pi\quad, 
\label{eq:3.7e}
\ee
and
\be
J^{\rm (1,s)} = -3J^{\rm (1,t)} = -24 J^{(1)}/\pi\quad. 
\label{eq:3.7f}
\ee
\eml%
This relation between the bare values of $J^{\rm (1,s)}$ and $J^{\rm (1,t)}$ 
will be important later. Notice that $J^{\rm (1,s)}<0$, while
$J^{\rm (1,t)}>0$, $J^{\rm (2,t)}>0$.

Comparing these expression to the correspoding ones for an electron-electron 
interaction,\cite{us_fermions} one sees that they have the same structure
except for the frequency sector. Transforming from Matsubara frequency
space into time space reveals that the annealed disorder corresponds to an
interaction that is infinitely long-ranged in time. This is physically
plausible, as has been explained in Sec.\ \ref{subsec:II.B}.

\subsection{Renormalizability considerations}
\label{subsec:III.C}

For reasons explained in Appendices \ref{app:A} and \ref{app:B}, 
we will choose a field-theoretic
RG method\cite{ZJ} over a momentum-shell RG.\cite{WilsonKogut} 
Before we start analyzing our model by 
means of this method, we need to ask whether the model is renormalizable, 
and how many renormalization constants are required.
Much is known about the renormalization properties of the NL$\sigma$M,
Eq.\ (\ref{eq:3.1}), with additional {\em instantaneous} interaction terms. 
The pure NL$\sigma$M is known to be renormalizable with two renormalization
constants, one for the coupling constant $G$ and one field renormalization
constant.\cite{ZJ} The frequency coupling $H^{(1)}$ turns out to not carry
a renormalization constant of its own. In the presence of an instantaneous
interaction, the proof of renormalizability for the NL$\sigma$M breaks
down, and the renormalizability of the model has never been proven. However,
there is much
evidence that the model is still renormalizable, with two additional
renormalization constants for the interaction, and with $H^{(1)}$ acquiring
a renormalization constant of its own. The two renormalization constants
for the interaction terms correspond to symmetric and antisymmetric
combinations of terms bilinear in $Q$, respectively.\cite{Pruisken,us_NPB} 
The same arguments apply to the present model, and are given in
Appendix \ref{app:B}. From Eqs.\ (\ref{eqs:B.1}), 
we conclude that we need to write
${\cal A}_{\rm ann}^{(1,s)} = {\cal A}_{+}^{(1,s)} +
{\cal A}_{-}^{(1,s)}$, and analogously split
${\cal A}_{\rm ann}^{(1,t)}$ and ${\cal A}_{\rm ann}^{(2,t)}$, with
\bml
\label{eqs:3.8}
\bea
{\cal A}_{+}^{\rm (1,s)}&=&-2\pi T J^{\rm (1,s)}_{+}\int d{\bf x}
  \sum_{nm}\sum_{\alpha}\sum_{r=0,3}\Bigl[{^0_rQ}_{nm}^{\alpha\alpha}({\bf x})
\nonumber\\
&&\times {^0_rQ}_{nm}^{\alpha\alpha}({\bf x})
   + \frac{1}{2}\sum_i {^i_rQ}_{nn}^{\alpha\alpha}({\bf x})\,
                       {^i_rQ}_{mm}^{\alpha\alpha}({\bf x})\Bigr]\quad,
\nonumber\\
\label{eq:3.8a}\\
{\cal A}_{-}^{\rm (1,s)}&=&-2\pi T J^{\rm (1,s)}_{-}\int d{\bf x}
  \sum_{nm}\sum_{\alpha}\sum_{r=0,3}\Bigl[{^0_rQ}_{nm}^{\alpha\alpha}({\bf x})
\nonumber\\
&&\times {^0_rQ}_{nm}^{\alpha\alpha}({\bf x})
   - \frac{1}{2}\sum_i {^i_rQ}_{nn}^{\alpha\alpha}({\bf x})\,
                       {^i_rQ}_{mm}^{\alpha\alpha}({\bf x})\Bigr]\quad,
\nonumber\\
\label{eq:3.8b}\\
{\cal A}_{+}^{\rm (1,t)}&=&-2\pi T J^{\rm (1,t)}_{+}\int d{\bf x}
  \sum_{nm}\sum_{\alpha}\sum_{r=0,3}\Bigl[\sum_{i=1}^{3}
        {^i_rQ}_{nm}^{\alpha\alpha}({\bf x})
\nonumber\\
&&\hskip -8pt \times {^i_rQ}_{nm}^{\alpha\alpha}({\bf x})
   + \frac{1}{2}\sum_i\left({{3\atop -}\atop{-\atop -}}\right)_i
      {^i_rQ}_{nn}^{\alpha\alpha}({\bf x})\,
                       {^i_rQ}_{mm}^{\alpha\alpha}({\bf x})\Bigr]\quad,
\nonumber\\
\label{eq:3.8c}\\
{\cal A}_{-}^{\rm (1,t)}&=&-2\pi T J^{\rm (1,t)}_{-}\int d{\bf x}
  \sum_{nm}\sum_{\alpha}\sum_{r=0,3}\Bigl[\sum_{i=1}^{3}
        {^i_rQ}_{nm}^{\alpha\alpha}({\bf x})
\nonumber\\
&&\hskip -8pt\times {^i_rQ}_{nm}^{\alpha\alpha}({\bf x})
   - \frac{1}{2}\sum_i\left({{3\atop -}\atop{-\atop -}}\right)_i
      {^i_rQ}_{nn}^{\alpha\alpha}({\bf x})\,
                       {^i_rQ}_{mm}^{\alpha\alpha}({\bf x})\Bigr]\quad,
\nonumber\\
\label{eq:3.8d}\\
{\cal A}_{+}^{\rm (2,t)}&=&2\pi T J^{\rm (2,t)}_{+}\int d{\bf x}
  \sum_{nm}\sum_{\alpha}\sum_{r=0,3}\Bigl[\sum_{i=1}^{3}
        {^i_rQ}_{nn}^{\alpha\alpha}({\bf x})
\nonumber\\
&&\hskip -8pt\times {^i_rQ}_{mm}^{\alpha\alpha}({\bf x})
   + \frac{1}{2}\sum_i\left({{3\atop -}\atop{-\atop -}}\right)_i
      {^i_rQ}_{nm}^{\alpha\alpha}({\bf x})\,
                       {^i_rQ}_{nm}^{\alpha\alpha}({\bf x})\Bigr]\quad,
\nonumber\\
\label{eq:3.8e}\\
{\cal A}_{-}^{\rm (2,t)}&=&2\pi T J^{\rm (2,t)}_{-}\int d{\bf x}
  \sum_{nm}\sum_{\alpha}\sum_{r=0,3}\Bigl[\sum_{i=1}^{3}
        {^i_rQ}_{nn}^{\alpha\alpha}({\bf x})
\nonumber\\
&&\hskip -8pt\times {^i_rQ}_{mm}^{\alpha\alpha}({\bf x})
   - \frac{1}{2}\sum_i\left({{3\atop -}\atop{-\atop -}}\right)_i
     {^i_rQ}_{nm}^{\alpha\alpha}({\bf x})\,
                       {^i_rQ}_{nm}^{\alpha\alpha}({\bf x})\Bigr]\quad,
\nonumber\\
\label{eq:3.8f}
\eea
\eml%
In writing Eqs.\ (\ref{eqs:3.8}) we have made use of Eqs.\ (\ref{eq:3.2a}) -
(\ref{eq:3.2c}). The symbol $\left({{3\atop -}\atop{-\atop -}}\right)_i$
is a shorthand for $3\delta_{i0} - \sum_{j=1}^{3}\delta_{ij}$.
The $J^{(1,s)}_{\pm}$ are coupling constants whose bare values are equal,
\bml
\label{eqs:3.9}
\be
J^{(1,s)}_{+} = J^{(1,s)}_{-} = J^{(1,s)}\quad,
\label{eq:3.9a}
\ee
but in general they renormalize differently. Similarly,
\bea
J^{(1,t)}_{+} = J^{(1,t)}_{-} = J^{(1,t)}\quad,
\label{eq:3.9b}\\
J^{(2,t)}_{+} = J^{(2,t)}_{-} = J^{(2,t)}\quad,
\label{eq:3.9c}
\eea
\eml%
in the bare theory, but under renormalization these equalities do not
in general remain valid.
All of the $J_{+}$ require only one renormalization constant, which we will
denote by $Z_{+}$, and the $J_{-}$ require another one, $Z_{-}$. In addition,
a renormalization constant for $H^{(2)}$ is needed. 

In addition to the relations given by Eqs.\ (\ref{eqs:3.9}), there is the
relation between $J^{(1,s)}$ and $J^{(1,t)}$ given by Eq.\ (\ref{eq:3.7f}).
It will turn out that these constraints leads to a degeneracy in the
RG flow. This is most easily handled by relaxing the condition, 
Eq.\ (\ref{eq:3.9a}). Instead of Eqs.\ (\ref{eqs:3.9}) and (\ref{eq:3.7f})
we therefore write
\bml
\label{eqs:3.10}
\bea
J^{\rm (1,s)}_{\pm} &=& J^{\rm (1,s)} \pm \Delta\quad,
\label{eq:3.10a}\\
J^{\rm (1,t)}_{\pm} &=& J^{\rm (1,t)}\quad,
\label{eq:3.10b}\\
J^{\rm (2,t)}_{\pm} &=& J^{\rm (2,t)}\quad,
\label{eq:3.10c}
\eea
and
\be
J^{\rm (1,s)} + 3J^{\rm (1,t)} = 0\quad.
\label{eq:3.10d}
\ee
\eml%
Choosing $\Delta \neq 0$ will remove the degeneracy in the RG flow. In
the end, we will consider the limit $\Delta\rightarrow 0$ to obtain the
behavior of our original model.

\section{Renormalization to one-loop order}
\label{sec:IV}

\subsection{Perturbation theory}
\label{subsec:IV.A}

\subsubsection{Gaussian propagators}
\label{subsubsec:IV.A.1}

We now perform a one-loop RG analysis of the model defined in 
Sec.\ \ref{sec:III}. To this end, we expand the action in powers of
the matrix $q$ defined by Eq.\ (\ref{eq:3.2e}). To Gaussian order
we find
\bml
\label{eqs:4.0}
\be
{\cal A} = \frac{-4}{G}\,\frac{1}{V}\sum_{\bf p}\sum_{12}\sum_{i,r}
           {^i_rq}_{12}({\bf p})\,\Gamma^{(2)}({\bf p},\Omega_{n_1-n_2})
           \,{^i_rq}_{12}(-{\bf p})\ ,
\label{eq:4.0a}
\ee
with
\bea
\Gamma^{(2)}({\bf k},\Omega_n) &=& {\bf k}^2/G + H^{(1)}\Omega_n + H^{(2)}2\pi T
\nonumber\\
&&   + \delta_{\alpha_1\alpha_2}\pi TG\,\left[
       \delta_{i0} J_{\rm s} + (1-\delta_{i0}) J_{\rm t}\right]\quad.
\label{eq:4.0b}
\eea
\eml%
the bare two-point vertex.
The Gaussian $q$-propagators are obtained by inverting this quadratic form. 
We find
\bml
\label{eqs:4.1}
\be
\langle {^i_rq}_{12}({\bf k})\,{^j_sq}_{34}({\bf p})\rangle = 
   \delta_{{\bf k},-{\bf p}}\,\delta_{13}\,\delta_{24}\,\delta_{rs}\,
   \delta_{ij}\,\frac{G}{8}\,{^i{\cal D}}_{12}({\bf k})\quad.
\label{eq:4.1a}
\ee
Here $\langle\ldots\rangle$ denotes a Gaussian average, and
$1\equiv (n_1,\alpha_1)$, etc., are indices that comprise both the
Matsubara frequency index and the replica label. The propagators
$^i{\cal D}$ read
\bea
{^0{\cal D}}_{12}({\bf k}) = {\cal D}_{n_1-n_2}({\bf k}) 
   + \delta_{\alpha_1\alpha_2}\ \Delta{\cal D}^{\rm s}_{n_1-n_2}({\bf k})\quad,
\label{eq:4.1b}\\
{^{1,3.3}{\cal D}}_{12}({\bf k}) = {\cal D}_{n_1-n_2}({\bf k}) 
   + \delta_{\alpha_1\alpha_2}\ \Delta{\cal D}^{\rm t}_{n_1-n_2}({\bf k})\quad,
\label{eq:4.1c}
\eea
where
\bea
{\cal D}_n({\bf k}) &=& \frac{1}{{\bf k}^2 + GH^{(1)}\Omega_n + GH^{(2)}2\pi T}
                       \quad,
\label{eq:4.1d}\\
{\cal D}^{\rm s,t}_n({\bf k}) &=& \frac{1}{{\bf k}^2 + GH^{(1)}\Omega_n
   + GH^{(2)}2\pi T + GJ_{\rm s,t} 2\pi T}\quad,
\nonumber\\
\label{eq:4.1e}
\eea
with
\be
\Delta{\cal D}_n^{\rm s,t}({\bf k}) = {\cal D}_n^{\rm s,t}({\bf k})
    - {\cal D}_n({\bf k})\quad,
\label{eq:4.1f}
\ee
and
\bea
J_{\rm s} = \frac{1}{2}\left(J^{\rm (1,s)}_{+} + J^{\rm (1,s)}_{-}\right)
            - \frac{3}{4}\left(J^{\rm (2,t)}_{+} - J^{\rm (2,t)}_{-}\right)
            \quad,
\label{eq:4.1g}\\
J_{\rm t} = \frac{1}{2}\left(J^{\rm (1,t)}_{+} + J^{\rm (1,t)}_{-}\right)
            + \frac{1}{4}\left(J^{\rm (2,t)}_{+} - J^{\rm (2,t)}_{-}\right)
            \quad.
\label{eq:4.1h}
\eea
\eml%

\subsubsection{One-loop corrections}
\label{subsubsec:IV.A.2}

\begin{figure}[h,t]
\centerline{\psfig{figure=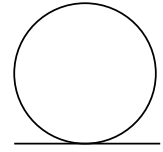,width=20mm}\vspace*{5mm}}
\caption{Structure of diagrams that renormalize the two-point vertex.}
\label{fig:1}
\end{figure}

By expanding the action to $O(q^4)$ and calculating all diagrams with
the topological structure shown in Fig.\ \ref{fig:1}, we obtain the
one-loop corrections $\delta G$, $\delta H^{(1)}$, etc. to the coupling
constants in the Gaussian propagators, Eqs.\ (\ref{eqs:4.1}), or
the 2-point vertex, Eq.\ (\ref{eq:4.0b}). The explicit calculation is
similar to the one for the case of an instantaneous 
interaction,\cite{us_R} but substantially simpler due to the absence 
of cubic terms in the $q$-expansion. We find
\bml
\label{eqs:4.3}
\bea
\delta G &=& \frac{G^2}{16}\,\left(K_{+} + K_{-}\right)\,I_2\quad,
\label{eq:4.3a}\\
\delta H^{(1)} &=& \frac{-GH^{(1)}}{16}\,\left(K_{+} + K_{-}\right)\,I_2\quad,
\label{eq:4.3b}\\
\delta H^{(2)} &=& \frac{-G}{16}\,\left[H^{(2)}\left(K_{+} + K_{-}\right)
   + \frac{3}{2}\,\left(J^{\rm (1,t)}_{+} + J^{\rm (1,t)}_{-}\right.\right.
\nonumber\\
&&\left.\left.\hskip -34pt + \frac{1}{2}\,J^{\rm (2,t)}_{+} 
     - \frac{1}{2}\,J^{\rm (2,t)}_{-}\right)
   \left(L_{+} + L_{-} - 2J^{\rm (2,t)}_{+} + 2J^{\rm (2,t)}_{-}\right)
   \right]I_2
\nonumber\\
&&-\frac{3G}{16}\,\left(J^{\rm (2,t)}_{+} + J^{\rm (2,t)}_{-}
   -\frac{1}{2}\,L_{+} + \frac{1}{2}\,L_{-}\right)\,I_1
\nonumber\\
&&+ \frac{G}{32}\,\left(K_{+} - K_{-}\right)\,I_1\quad,
\label{eq:4.3c}\\
\delta J_{\rm s} &=& \frac{-G}{8}\,\left(J_{\rm s}^2 + 3J_{\rm t}^2\right)\,I_2
   + \frac{3G}{16}\,\left(J^{\rm (2,t)}_{+} + J^{\rm (2,t)}_{-}\right.
\nonumber\\
&&\left.  -\frac{1}{2}\,L_{+} + \frac{1}{2}\,L_{-}\right)\,I_1\quad,
\label{eq:4.3d}\\
\delta J_{\rm t} &=& \frac{-G}{16}\,\left(J^{\rm (1,t)}_{+} + J^{\rm (1,t)}_{-}
   + \frac{1}{2}\,J^{\rm (2,t)}_{+} - \frac{1}{2}\,J^{\rm (2,t)}_{-}\right)
\nonumber\\
&&\hskip -22pt\times \left(J^{\rm (1,s)}_{+} + J^{\rm (1,t)}_{+}
   + J^{\rm (1,s)}_{-} + J^{\rm (1,t)}_{-} - J^{\rm (2,t)}_{+} 
        + J^{\rm (2,t)}_{-}\right)I_2
\nonumber\\
&&- \frac{G}{16}\,\left(J^{\rm (2,t)}_{+} + J^{\rm (2,t)}_{-}
      - \frac{1}{2}\,L_{+} + \frac{1}{2}\,L_{-}\right)\,I_1\quad,
\nonumber\\
\label{eq:4.3e}
\eea
\eml%
Here we have defined linear combinations of coupling constants,
\bml
\label{eqs:4.4}
\bea
K_{\pm} &=& J^{\rm (1,s)}_{\pm} + 3J^{\rm (1,t)}_{\pm} = \pm\Delta\quad,
\label{eq:4.4a}\\
L_{\pm} &=& J^{\rm (1,s)}_{\pm} - J^{\rm (1,t)}_{\pm}\quad,
\label{eq:4.4b}
\eea
\eml%
where the second equality in Eq.\ (\ref{eq:4.4a}) is due to 
Eqs.\ (\ref{eqs:3.10}). This will be important later. 
We have also defined one-loop integrals
\bml
\label{eqs:4.5}
\bea
I_1 &=& G\int d{\bf p}\ {\cal D}_n({\bf p}) = -{\bar G}/G\epsilon\quad,
\label{eq:4.5a}\\
I_2 &=& G\int d{\bf p}\ 2\pi T \sum_n \left({\cal D}_n({\bf p})\right)^2
    = -{\bar G}/GH^{(1)}\epsilon\quad.
\label{eq:4.5b}
\eea
\eml%
Here $\epsilon = d-2$, and ${\bar G} = GS_d/(2\pi)^d$ with $S_d$ the
surface area of the $(d-1)$-sphere. In giving the second equalities in
Eqs.\ (\ref{eqs:4.5}) we have chosen to use dimensional regularization,
and in what follows we will use a field-theoretic RG method. At a
perturbative level, this is a matter of choice, and we could just as
well use the momentum-shell RG method. In that case, the factors of
$-1/\epsilon$ in Eqs.\ (\ref{eqs:4.5}) would be replaced by $\ln b$,
with $b$ the RG length rescaling factor. For arguments that go beyond
perturbation theory, however, it is advantageous to use the field theory
approach, as is explained in the Appendices \ref{app:A} and \ref{app:B}.

In addition to these renormalizations of the two-point propagator or
vertex function, we will also need the one-point vertex $\Gamma^{(1)}$
to one-loop order. This is given by the diagram shown in
\begin{figure}[h,t]
\centerline{\psfig{figure=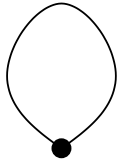,width=15mm}\vspace*{5mm}}
\caption{Structure of diagrams that renormalize the one-point vertex.}
\label{fig:2}
\end{figure}
Fig.\ \ref{fig:2}, and a simple calculation yields
\be
\Gamma^{(1)} \equiv \langle{^0_0Q}_{nn}^{\alpha\alpha}({\bf x})\rangle^{-1}
   = 1 - \frac{G}{16}\,\left(K_{+} + K_{-}\right)\,I_2\quad.
\label{eq:4.6}
\ee

For later reference, we notice that the one-loop corrections to $G$,
$H^{(1)}$, and $\Gamma^{(1)}$ vanish in the limit $\Delta\rightarrow 0$,
and that 
\be
\delta J_{\rm s} + 3\delta J_{\rm t} = 0\quad,
\label{eq:4.7}
\ee
as can be seen by using Eqs.\ (\ref{eqs:3.10}). Furthermore, a calculation
shows that
\be
\delta H^{(1)} + \delta H^{(2)} + \delta J_{\rm s} = \frac{G}{16}\,\Delta\,I_1
  \quad,
\label{eq:4.8}
\ee
which also vanishes as $\Delta\rightarrow 0$.

\subsection{Renormalization}
\label{subsec:IV.B}

\subsubsection{Renormalization constants}
\label{subsubsec:IV.B.1}

We now proceed to renormalize the theory, i.e., we absorb the
singularities in the $\epsilon\rightarrow 0$ limit that are
present in perturbation theory into renormalization constants.
We define renormalized coupling constants $g$, $h^{(1)}$, etc.,
by
\bea
{\bar G} &=& \mu^{-\epsilon} Z_g g\ \,,\ \,\, 
H^{(1)} = Z_{h}^{(1)}h^{(1)}\ ,\ 
H^{(2)} = Z_{h}^{(2)}h^{(2)}\ ,
\nonumber\\
J^{\rm (1,s)}_{+} &=& Z_{+}j^{\rm (1,s)}_{+}\ ,\ 
J^{\rm (1,t)}_{+} = Z_{+}j^{\rm (1,t)}_{+}\ ,\ 
J^{\rm (2,t)}_{+} = Z_{+}j^{\rm (2,t)}_{+}\ , 
\nonumber\\
J^{\rm (1,s)}_{-} &=& Z_{-}j^{\rm (1,s)}_{-}\ ,\ 
J^{\rm (1,t)}_{-} = Z_{-}j^{\rm (1,t)}_{-}\ ,\ 
J^{\rm (2,t)}_{-} = Z_{-}j^{\rm (2,t)}_{-}\ ,
\nonumber\\ 
\label{eq:4.9}
\eea
where $\mu$ is an arbitrary momentum scale. The renormalization
statement is\cite{ZJ}
\bea
\Gamma_{\rm R}^{(N)}({\bf p},\Omega_n;g,h,j_{+},j_{-};\mu) &=&
\nonumber\\
&&\hskip -65pt
   Z^{(N/2)}\Gamma^{(N)}({\bf p},\Omega_n;G,H,J_{+},J_{-})\quad.
\label{eq:4.10}
\eea
Here $\Gamma_{\rm R}^{(N)}$ is the renormalized $N$-point vertex
function, $Z$ is the field renormalization constant, and $H$
and $J_{\pm}$ represent the various frequency and annealed disorder
coupling constants. The assertion that {\em all} vertex functions
can be made finite to all orders in the loop expansion by the five
renormalization constants defined in Eq.\ (\ref{eqs:4.1}), plus the
field renormalization constant, is equivalent to saying that the
theory is renormalizable with these renormalization constants. 
As we have mentioned before, there is strong
evidence for this statement to be true, which is recapitulated in 
Appendix \ref{app:B}, but it has not been rigorously proven.

Assuming that the theory is renormalizable, the six equations,
Eqs.\ (\ref{eqs:4.3}) and (\ref{eq:4.6}), suffice to determine the
six renormalization constants to one-loop order. While it is
possible to do so for arbitrary bare values of the coupling
constants, the results simplify substantially if one uses
Eqs.\ (\ref{eqs:3.10}). Using minimal subtraction,\cite{ZJ} and taking
the limit $\Delta\rightarrow 0$, we obtain
\bml
\label{eqs:4.11}
\bea
Z &=& 1 + O(g^2)\quad,
\label{eq:4.11a}\\
Z_g &=& 1 + O(g^2)\quad,
\label{eq:4.11b}\\
Z_h^{(1)} &=& 1 + O(g^2)\quad,
\label{eq:4.11c}\\
Z_h^{(2)} &=& 1 + \frac{g}{\epsilon}\,\kappa(g,h,j_{+},j_{-})/h^{(2)}\quad,
\label{eq:4.11d}\\
Z_{+} &=& 1 + \frac{g}{\epsilon}\,\frac{2\phi_{\rm s}(g,h,j_{+},j_{-})}
          {j^{\rm (1,s)}_{+} + j^{\rm (1,s)}_{-}} + O(g^2)\quad,
\label{eq:4.11e}\\
Z_{-} &=& 1 + \frac{g}{\epsilon}\,\frac{2\phi_{\rm s}(g,h,j_{+},j_{-})}
          {j^{\rm (1,s)}_{+} + j^{\rm (1,s)}_{-}} + O(g^2)\quad.
\label{eq:4.11f}
\eea
\eml%
Here $\kappa$ and $\phi_{\rm s,t}$ are functions of the renormalized
coupling constants that are given by $\delta H^{(2)}$ and
$\delta J_{\rm s,t}$ as functions of the bare ones,
\bml
\label{eqs:4.12}
\bea
\kappa(G,H,J) &=& -\epsilon\,\delta H^{(2)}(G,H,J)/G\quad,
\label{eq:4.12a}\\
\phi_{\rm s,t}(G,H,J) &=& -\epsilon\,\delta J_{\rm s,t}(G,H,J)/G\quad.
\label{eq:4.12b}
\eea
An inspection shows that, in the limit $\Delta\rightarrow 0$,
\be
\kappa(G,H,J) = -\phi_{\rm s,t}(G,H,J)\quad.
\label{eq:4.12c}
\ee
\eml%
Notice that $Z_+ = Z_-$, at least to one-loop order. Since the bare values
of the various $J_{\pm}$ are identical, this means that the renormalized
values are also identical, and we can drop the distinction between the
$j_+$ and the $j_-$. We will thus write $j^{\rm (1,s)}_{+} = j^{\rm (1,s)}_{-}
\equiv j^{\rm (1,s)}$, etc. We note that this is a consequence of the
relations expressed by Eqs.\ (\ref{eqs:3.10}), and would not necessarily 
be true for more general models.

\subsubsection{Flow equations and their solutions}
\label{subsubsec:IV.B.2}

We now are in a position to determine the RG flow equations for the
coupling constants. Defining $\ell = -\ln\mu$ (or $\ell = \ln b$ in
an alternative momentum-shell approach), and using Eq.\ (\ref{eq:4.12c}), 
we obtain from Eqs.\ (\ref{eq:4.9}) and (\ref{eqs:4.11}), 
\bml
\label{eqs:4.13}
\bea
\frac{dg}{d\ell} &=& -\epsilon g + O(g^3)\quad,
\label{eq:4.13a}\\
\frac{dh^{(1)}}{d\ell} &=& O(g^2)\quad,
\label{eq:4.13b}\\
\frac{dh^{(2)}}{d\ell} &=& -g\,\phi_{\rm s}(g,h,j) + O(g^2)\quad,
\label{eq:4.13c}\\
\frac{dj^{(1,t)}}{d\ell} &=& \frac{-g}{3}\,\phi_{\rm s}(g,h,j) + O(g^2)\quad.
\label{eq:4.13d}
\eea
The flow of the remaining coupling constants $j$ can be obtained by
relating them to $j^{(1,t)}$. This is a consequence of there
being only two renormalization constants for all of the $J$. We obtain
\bea
j^{(1,s)}&=&j^{(1,t)}\,J^{(1,s)}/J^{(1,t)}
                = -3\,j^{(1,t)}\quad,
\label{eq:4.13e}\\
j^{(2,t)}&=&
                \frac{J^{(2,t)}}{J^{(1,t)}}\,j^{(1,t)}\quad. 
\label{eq:4.13f}
\eea 
\eml%

In order to determine the nature of these flows, we calculate $\phi_{\rm s}$
from Eqs.\ (\ref{eq:4.3d}) and (\ref{eq:4.12b}). We find
\be
\phi_{\rm s}(g,h,j) = \frac{-3}{2}\,\frac{(j^{(1,t)})^2}{h^{(1)}}\,\left[
   1 - \frac{j^{(2,t)}h^{(1)}}{4(j^{(1,t)})^2}\right] + O(g)\quad.
\label{eq:4.14}
\ee
We see that $\phi_{\rm s}<0$, unless $J^{(2,t)}$ is larger than $(J^{(1,t)})^2$
in suitable units (note that the $J$'s and $H$'s all have the dimensions of
a density of states). This makes physical sense: From Eqs.\ (\ref{eq:3.7c})
and (\ref{eq:3.7d}) we see that ${\cal A}^{(1,t)}$ and ${\cal A}^{(2,t)}$
are spin-triplet interactions with different signs. $J^{(1,t)}>0$ promotes
ferromagnetism, and $J^{(2,t)}>0$ weakens that tendency. In two-dimensions,
for physically
sensible values of the coupling constants, we thus have $\phi_{\rm s}<0$,
and $h^{(2)}$ and $j^{(1,t)}$ both scale to infinity. In $d>2$, the RG
flow equations can be solved explicitly and shown to describe a quantum
phase transition by introducing, as in Ref.\ \onlinecite{us_IFS},
a scaling variable $y = gj^{(1,t)}/h^{(1)}$ that obeys
\be
\frac{dy}{d\ell} = -\epsilon y + y^2/2 + O(y^3)\quad.
\label{eq:4.15}
\ee
We see that Eq.\ (\ref{eq:4.15}) allows for a fixed point value 
$y^* = \epsilon + O(\epsilon^2)$. Denoting the deviation from this
fixed point value by $\delta y$, we find
\bml
\label{eqs:4.16}
\be
\delta y(b) = \delta y(b=1)\,b^{\,\epsilon + O(\epsilon^2)}\quad,
\label{eq:4.16a}
\ee
and
\bea
h^{(2)}(b)&=&h^{(2)}(b=1)\,b^{\,\epsilon + O(\epsilon^2)}\quad,
\label{eq:4.16b}\\
j^{(1,t)}(b)&=&j^{(1,t)}(b=1)\,b^{\,\epsilon + O(\epsilon^2)}\quad,
\label{eq:4.16c}\\
h^{(1)}(b)&=&h^{(1)}(b=1)\,b^{0 + O(\epsilon^2)}\quad,
\label{eq:4.16d}\\
g(b)&=&g(b=1)\,b^{-\epsilon + O(\epsilon^2)}\quad.
\label{eq:4.16e}
\eea
\eml%
The behavior of all observables of interest can be deduced from the
above flows, see Sec.\ \ref{subsec:V.A} below.

\section{Discussion}
\label{sec:V}

\subsection{Physical interpretation, and results}
\label{subsec:V.A}

For a physical interpretation of our results we first need to
relate physical observables to the coupling constants of
our theory. Some observables can be identified directly in analogy to
the corresponding identification in the case of an instantaneous
electron-electron interaction. From the derivation of the
NL$\sigma$M, $G$ is known to be related to the bare conductivity
$\sigma$ via\cite{Wegner,us_R}
\bml
\label{eqs:5.1}
\be
\sigma = 8/\pi G\quad.
\label{eq:5.1a}
\ee
The single-particle or tunneling density of states $N$ at an energy
$\omega$ from the Fermi level is related to the one-point vertex
by\cite{F}
\be
N(\epsilon_{\rm F} + \omega) = \frac{4}{\pi}\,\left(\Gamma^{(1)}\right)^{-1}
   (i\omega_n\rightarrow \omega + i0)\quad.
\label{eq:5.1b}
\ee
\eml%
Equations (\ref{eq:4.13a}) and (\ref{eq:4.11a}) show that $\sigma$ and $N$
are not renormalized, at least to one-loop order,
\bml
\label{eqs:5.2}
\bea
\frac{d\sigma}{d\ell}&=&O(g^2)\quad,
\label{eq:5.2a}\\
\frac{dN}{d\ell}&=&O(g^2)\quad.
\label{eq:5.2b}
\eea
\eml%
The scaling behavior of the relevant operator $\delta y$, Eq.\ (\ref{eq:4.16a}),
determines the correlation length exponent. Denoting the dimensionless
distance from the critical point by $t$, and the correlation length by $\xi$,
one finds for small $t$
\bml
\label{eqs:5.3}
\be
\xi \propto \vert t\vert^{-\nu}\quad,
\label{eq:5.3a}
\ee
with a correlation length exponent
\be
\nu = 1/\epsilon + O(1)\quad.
\label{eq:5.3b}
\ee
\eml%

Other quantities of interest are various susceptibilities, in particular
the specific heat coefficient $\gamma_V = C_V/T$, the spin susceptibility
$\chi_{\rm s}$, and the density susceptibility $\partial n/\partial\mu$.
Their relations to the coupling constants in the field theory are less
obvious. We therefore use scaling arguments, in conjunction with
perturbation theory for the free energy, to determined their respective
cricital behavior. We start with a homogeneity law for the free energy.
From the Gaussian propagators, Eqs.\ (\ref{eqs:4.1}), we see that that
in principle there are three different time scales in the theory,
given by
\bml
\label{eqs:5.4}
\bea
\tau_1&=&\xi^d gh^{(1)} \sim \xi^{z_1}\quad,
\label{eq:5.4a}\\
\tau_2&=&\xi^d gh^{(2)} \sim \xi^{z_2}\quad,
\label{eq:5.4b}\\
\tau_3&=&\xi^d gj^{(1,t)} \sim \xi^{z_3}\quad,
\label{eq:5.4c}
\eea
\eml%
Here $z_{1,2,3}$ are the dynamical exponents related to these time scales.
To one-loop order we have
\bml
\label{eqs:5.4'}
\bea
z_1 &=& d - \epsilon + O(\epsilon^2) = 2 + O(\epsilon^2)\quad,
\label{eq:5.4'a}\\
z_2 &=& z_3 = d - \epsilon + \epsilon + O(\epsilon^2) = d + O(\epsilon^2)\quad,
\label{eq:5.4'b}
\eea
\eml%
leaving us with two times scales
and dynamical exponents. The free energy density $f$ therefore has two
different scaling parts, and we can write
\bea
f(t,T,\ldots) &=& b^{-(d+z_1)}\,f_1(t\,b^{1/\nu},T\,b^{z_1},T\,b^{z_2},\ldots)
\nonumber\\
&& + b^{-(d+z_2)}\,f_2(t\,b^{1/\nu},T\,b^{z_1},T\,b^{z_2},\ldots)\ .
\label{eq:5.5}
\eea
Here $f_1$ and $f_2$ are scaling functions, and the ellipses denote the
dependence of $f$ on external fields that are not shown explicitly.

The specific heat coefficient is obtained by differenting $f$ twice
with respect to $T$. The leading contribution is obtained by differentiating
$f_1$ with respect to the temperature scale that carries the dynamical
exponent $z_2$. This yields
\bml
\label{eqs:5.6}
\be
\gamma_V(t) \sim \vert t\vert^{-\alpha}\quad,
\label{eq:5.6a}
\ee
with a critical exponent
\be
\alpha = \nu (2z_2 - d - z_1) = 1 + O(\epsilon)\quad.
\label{eq:5.6b}
\ee
\eml%
To ascertain that this leading contribution has a nonzero prefactor
we check against perturbation theory for the free energy, which is
given in Appendix\ \ref{app:D}. From Eqs.\ (\ref{eq:D.1b},\ref{eq:D.1c}) 
we see that there is indeed a contribution from differentiating twice
with respect to the temperature in the propagators, which carries a
dynamical exponent $z_2$. The temperature prefactor in the expression
$f=-(T/V)\ln Z$ for the free energy density has been absorbed into
the frequency integration measure. The frequency, however, scales like
a wavenumber squared, and therefore carries an exponent $z_1$.

A very similar argument applies to the spin susceptibility. A
magnetic field $B$ couples to the electrons via a Zeeman term (amongst
other coupling mechanisms), and hence can scale like an energy or
temperature. The spin susceptibility is obtained by differentiating
$f$ twice with respect to $B$, and once therefore expects $\chi_{\rm s}$
to scale like the specific heat coefficient, viz.
\bml
\label{eqs:5.7}
\be
\chi_{\rm s}(t) \sim \vert t\vert^{-\gamma}\quad,
\label{eq:5.7a}
\ee
with a critical exponent
\be
\gamma = \alpha = 1 + O(\epsilon)\quad.
\label{eq:5.7b}
\ee
\eml%
Again, perturbation theory confirms that the leading contribution obtained
in this way is nonzero. This is easily seen from 
Eqs.\ (\ref{eq:D.1b},\ref{eq:D.1c}) by
taking into account that $B\neq 0$ leads to a mass $\mu_{\rm B}B$ in two of
the spin-triplet propagators that contribute to the Gaussian approximation
for the free energy.

Finally, we consider $\partial n/\partial\mu$. Although the chemical
potential $\mu$ is dimensionally an energy, it differs fundamentally
from either $T$ or $\mu_{\rm B}B$, since it represents the microscopic
energy or inverse time scale. As such, it must have an effective scale
dimension of zero. Consequently, we obtain from Eq.\ (\ref{eq:5.5}),
by differentiating twice with respect to $\mu$,
\be
(\partial n/\partial\mu)(t) = {\rm const.} + O(t^{\nu (d+z_1)})\quad.
\label{eq:5.8}
\ee
$\partial n/\partial\mu$ thus has only a weak nonanalytic $t$-dependence
in addition to a leading noncritical contribution. Again, this is
consistent with perturbation theory: The only $\mu$-dependence of
the free energy, Eq.\ (\ref{eq:D.1b}), is through the various coupling
constants in the propagators. All of these multiply either a frequency
or a temperature. Differentiation with respect to $\mu$ therefore does
not produce a singular integral unless $f$ itself becomes singular.
Power counting shows that this happens only for dimensions $d\leq -2$,
in agreement with Eq.\ (\ref{eq:5.8}). This failure of differentiation
with respect to a field to produce a singularity is an illustration of
a more general argument given in Ref.\ \onlinecite{us_local_OP}.

The physical interpretation of these results is now clear. The RG flow at
one-loop order is qualitatively the same as for electrons interacting via
an instantaneous interaction, see the comparison between the two flows
given in Appendix \ref{app:C}. In the latter case, the runaway flow of the
equivalent of $j^{\rm (1,t)}$ ($k_{\rm t}$ in Appendix \ref{app:C}) at
one-loop order in $d=2$ suggests a ferromagnetic ground state. In 
$d=2+\epsilon$ there is a phase transition where the 
homogeneous magnetic susceptibility diverges. This transition has been
identified with a ferromagnetic phase
transition where the magnetic susceptibility diverges like
$\chi_{\rm s}\sim \vert t\vert^{-\gamma}$, as in 
Eq.\ (\ref{eq:5.7a}).\cite{us_IFS,us_fm_local} 
The runaway flow thus simply reflects the fact
that $t$ is RG relevant at a ferromagnetic transition. The result of this 
interpretation agrees with a more direct, and more explicit,
theory for the ferromagnetic transition.\cite{us_fm_local} In the current
case, the theory describes an infinite-range version of this transition, 
due to the interaction being infinitely long-ranged in time. These 
considerations strongly suggest that the physical results we have derived 
above to one-loop order actually
hold to all orders in the loop expansion, as they do in the instantaneous
interaction case.\cite{us_IFS} In particular, we expect that 
$\partial n/\partial\mu$ is not renormalized to all orders, in agreement
with Ref.\ \onlinecite{us_local_OP}. It also follows that the phase
diagram for the present model is qualitatively similar to the one for
the interacting case, with a ferromagnetic transition {\em always}
preceding an MIT for $d\agt 2$, while for $d=3$ a direct transition
from a paramagnetic metal to a paramagnetic insulator is 
possible.\cite{us_IFS,us_R} There are, however,
differences in the detailed properties of the transition as compared to the
one studied in Refs.\ \onlinecite{us_fm_local} and \onlinecite{us_IFS}. For
instance, in the latter the specific heat has a much weaker singularity
than the spin susceptibility, while here they show the same scaling behavior.
In this respect the current case is reminiscient of the Brinkman-Rice
theory of the Hubbard MIT.\cite{BrinkmanRice}

Although the transition in the present model
is classical, in the sense that the order parameter is purely
static, it couples to quantum mechanical degrees of freedom in the form
of the diffusive electrons. An explicit description of the transition could
be obtained along the lines of Ref.\ \onlinecite{us_fm_local}. However,
given the schematic nature of our model, we will not pursue this here.
The same conclusion, namely that the model under consideration describes
a ferromagnetic transition of a classical nature, has recently been reached
by Vojta and Narayanan by means of very different arguments.\cite{TV_RN} 
We stress, however, that our goal here has not been to describe a magnetic 
transition. Rather, it was to resolve the conflict between the results of
Refs.\ \onlinecite{us_letter} and \onlinecite{us_local_OP}, and to check
whether or not our model of electrons with both quenched and annealed
disorder describes an unusual MIT. As we have seen, the answer to the latter
question is negative.

\subsection{Comparison with previous treatments}
\label{subsec:V.B}

The crucial difference between the treatment of the annealed disorder
model given above and the one in Ref.\ \onlinecite{us_letter} is
related to the occurrence of the coupling constant $H^{(2)}$. In
perturbation theory, i.e., in an expansion in powers of $q$, 
the annealed disorder generates terms that
have the structure of the last term on the right-hand side of
Eq.\ (\ref{eq:4.0b}), except that they are not constrained to
being diagonal in the replica index. There are two possible
interpretations of such terms. (1) They could represent 
terms quadratic in $Q$ that are not
diagonal in replica space. This was the interpretation given in
Ref.\ \onlinecite{us_letter}. (2) They could present a new term
{\em linear} in $Q$, which was not present in the original action.
The term with coupling constant $H^{(2)}$ introduced in the present
paper serves that purpose. By means of high-order perturbation theory
one could in principle distinguish between these two possibilities,
but this would be extremely cumbersome. Let us instead argue on
general structural and on physical grounds that the second 
interpretation is the correct one.

First, we have argued in
Sec.\ \ref{sec:II} that the annealed disorder, since it gets
averaged over at the level of the partition function, should indeed be
interpreted as an effective interaction between the electrons. As such,
all involved degrees of freedom must occur with the same replica
index, and the generation of an interaction term (i.e., one quadratic
in $Q$) for which this is not the case makes no physical sense.
Terms quadratic in $Q$ with more than one replica index are
characteristic for {\em quenched} disorder, and indeed the treatment
of the annealed disorder in Ref.\ \onlinecite{us_letter} was modeled
after that of quenched magnetic disorder. As we have argued above,
this is physically not plausible. 

Second, the appearance of a term with the structure of 
${\cal A}^{(2)}_{\Omega}$, Eq.\ (\ref{eq:3.4}), is plausible on
physical grounds. The term with coupling constant $H^{(1)}$ in
the NL$\sigma$M represents a frequency coupling with a microscopic
time scale, on the order of an inverse Fermi energy (in units where
$\hbar =1$). An interaction that is short-ranged in time does not
add a new time scale to the problem. It therefore renormalizes
$H^{(1)}$, but does not generate a new frequency coupling. An
interaction that is long-ranged in time, on the other hand,
does introduce a new time scale and hence a new frequency coupling.
In the general case of a frequency dependent interaction with a
continuum of time scales one would expect a frequency dependent
coupling constant $H$ whose scaling properties would have to be
studied by means of a functional RG. In our simple model where
the annealed disorder is static, which means that the resulting
effective interaction has an infinite range in time, one
additional frequency coupling suffices, which is $H^{(2)}$. 
The infinite time scale corresponds to a vanishing frequency scale,
in accord with the discontinuous frequency dependence $\sgn\Omega$ in 
Eq.\ (\ref{eq:3.4}). In this context, we note that the $H^{(2)}$
term does not represent an inelastic lifetime. Rather, it is a
true mass in the two-point propagators that is produced by the
long-ranged in time interaction. This is analogous to the mass
corresponding to the plasmon pole that is produced by an 
interaction that is long-ranged in space.

Third, the structure of the renormalization scheme used in
Ref.\ \onlinecite{us_letter} did not reflect the constraints
discussed in Appendix \ref{app:B}. This is only of minor concern if 
one neglects the coupling constant $J^{\rm (2,t)}$ and uses
one renormalization constant each for $J_{\rm s}$ and $J_{\rm t}$,
as was done in Ref.\ \onlinecite{us_letter}. It becomes crucial,
however, in the presence of $J^{\rm (2,t)}$, which forces the
issue of how many renormalization constants are needed.

Finally, the treatment of Ref.\ \onlinecite{us_letter} led to
results that were not consistent with independent, very general,
considerations. In particular, its prediction that 
$\partial n/\partial\mu$ is singularly renormalized, 
and critical at a MIT, contradicted one
of the results of Ref.\ \onlinecite{us_local_OP}. This point
requires some explanation. The critical behavior predicted implies
a nonanalytic dependence on the RG length scale, and hence a
nonanalytic dependence on the wavenumber $\vert{\bf q}\vert$ in
perturbation theory. In two-dimensions, this takes the form of
a $\ln\vert{\bf q}\vert$ term in perturbation theory that is
caused by the diffusive electron dynamics. In $d>2$, these same
integrals over diffusion poles lead to a $\vert{\bf q}\vert^{d-2}$
dependence.\cite{NLsM_footnote} 
The predicted critical behavior of $\partial n/\partial\mu$
at the MIT, and the mechanism that causes it, therefore implies a
nonanalytic wavenumber dependence of this susceptibility in the
metallic phase. However, it was shown on general grounds in
Ref.\ \onlinecite{us_local_OP} that $\partial n/\partial\mu$ is
an analytic function of the wavenumber for a large class of models,
which includes the one under consideration here. This
discrepancy prompted the current investigation, see the discussion 
in Sec.\ \ref{sec:I} above.

\subsection{Conclusion, and Outlook}
\label{subsec:V.C}

In conclusion, we have found that the treatment in Ref.\ \onlinecite{us_letter}
of the electron problem in the presence of annealed disorder, in addition to
quenched one,
was not correct. The perturbation theory was correct, but the assumptions made
about the RG structure of the theory were not. This was the reason for the
discrepancy between the explicit results found in Ref.\ \onlinecite{us_letter}
and later, more general considerations.\cite{us_local_OP} 
The current procedure,
which considers the annealed disorder as an effective electron-electron
interaction that is long-ranged in time, is physically and technically more
convincing. It yields results that are consistent with all of the available
information, and in particular with Ref.\ \onlinecite{us_local_OP}. Physically,
the model of static, annealed magnetic disorder representing a type of local
moments thus turns out to be less interesting than Ref.\ \onlinecite{us_letter}
had given reason to believe. Instead of describing an unusual MIT, the model
describes a variant of the ferromagnetic transition of itinerant electrons
that has been studied before. It is important to note that the same model
with quenched instead of annealed magnetic disorder is well known to contain
a MIT in $d=2+\epsilon$.\cite{F2} This serves to underscore the fundamental
physical difference between quenched and annealed disorder that we have
stressed several times in this paper. 

We finally mention a possible consequence of our observation, discussed in
Sec.\ \ref{subsec:V.B}, that an electron-electron interaction with more
than one time scale produces more than one frequency coupling in the
NL$\sigma$M, which in turn require additional renormalization constants.
(In the present case, there was one additional time scale, infinity,
one additional coupling, $H^{(2)}$, and
one additional renormalization constant.) At a MIT,
the coupling constant that was denoted above by $H^{(1)}$ acquires a
power-law frequency dependence. This is equivalent to saying that there
are infinitely many time scales in the problem, and this raises doubts about
the validity of renormalizing the action with just one renormalization constant
for the frequency coupling. It is therefore possible that a complete 
description of the dynamics near a MIT would require
a functional RG. A complete understanding of this problem would also require
a solution of the renormalizability problem for models of interacting
electrons that is explained in Appendix \ref{app:B}.

\acknowledgments
This work was initiated at the Aspen Center for Physics, and it is a pleasure
to thank the Center for its hospitality. This work was supported by
the NSF under grant Nos. DMR-98-70597, DMR-99-75259, DMR-01-32555, and
DMR-01-32726.

\appendix
\section{Momentum-shell versus field-theoretic renormalization}
\label{app:A}

In this appendix we motivate our choice of a field-theoretic formulation
of the renormalization procedure.

There exist two basic formulations of the RG, the
field-theoretic one that originated in high-energy physics,\cite{ZJ}
and Wilson's momentum-shell method,\cite{WilsonKogut} which was invented
for the study of critical points. After Wilson's breakthrough, it was
shown that the field-theoretic method can also be applied to critical
phenomena.\cite{ZJ} The relation between these two formulations of the
RG is complicated,\cite{Weinberg} but for our purposes we can restrict
ourselves to a few basic features.

In the Wilsonian method one renormalizes the Hamiltonian or action itself,
generating new interactions as one goes along, and checking all newly
generated terms for their scale dimensions, and hence for their being
RG relevant, irrelevant, or marginal. Irrelevant ones can be dropped,
while relevant or marginal ones must be added to the model and included
in a repetition of the renormalization process. In the field theoretic 
method, one renormalizes specific propagators or vertex functions, and 
one needs to know from the outset how many renormalization constants are 
needed in order to make all of the vertex functions finite to all orders.

For many models (e.g. for $\phi^4$-theory) there is only a small number
of relevant or marginal terms. In these cases, the momentum-shell method
is often preferred since it is physically more intuitive, and since it
provides an explicit check for the generation of additional terms that
must be kept. However, the NL$\sigma$M does not belong to this class,
as it has an infinite number of marginal terms in $d=2$: In an expansion
of Eq.\ (\ref{eq:3.1}) in powers of $q$, {\em all} terms are marginal.
It is a priori unclear how the infinitely many coupling constants
multiplying these terms renormalize, although their bare values all
coincide. The field-theoretic RG method proves that these
coupling constants all renormalize the same way.\cite{BrezinZJ,ZJ} This
fixes the structure of the renormalized theory, and it then suffices to 
consider a small number of vertex functions in order to determine the
renormalized theory explicitly. In the momentum-shell method, on the
other hand, one needs to explicitly consider a large number of vertex
functions or propagators (in principle infinitely many in the case of
the NL$\sigma$M) in order to do the same. 

The same considerations apply to the terms in addition to the NL$\sigma$M.
Equations (\ref{eqs:3.7}) add six coupling constants to the model.
Within a momentum-shell RG, one would have to consider $q^4$-vertices in
order to determine how they renormalize. The field-theoretic
method, on the other hand, allows us to argue that all of the $J$ split
into two pieces that pairwise renormalize in the same way, see
Sec.\ \ref{subsec:III.C} and Appendix\ \ref{app:B}. As a result, we need
to explicitly renormalize $q^2$-vertices only. This is the reason
why in this paper we choose the field-theoretic method over the 
momentum-shell one.

\section{Invariant decomposition of annealed disorder terms}
\label{app:B}

In this appendix we recall the answer to the following question: Consider
the NL$\sigma$M, Eq.\ (\ref{eq:3.1}), which is known to be renormalizable
in two-dimensions with two renormalization constants.\cite{BrezinZJ,ZJ}. 
Now add to this action symmetry breaking operators. How does this affect
the renormalizability, and how many additional renormalization constants
are needed?

For the case of operators that give some components of the basic field,
$Q({\bf x})$ in our case, a mass (massive insertions), this question has 
been studied in detail.\cite{soft_insertions} If the NL$\sigma$M is invariant 
under transformations that form a symmetry group ${\cal G}$, then the operators
in question must be expanded in a basis of irreducible representations of
${\cal G}$. All operators that belong to the same irreducible representation
renormalize the same way, i.e., for each irreducible representation one 
additional renormalization constant is needed.

In our case, it is most convenient to write the spin degrees of freedom
explicitly, and consider the complex numbers $Q^{\alpha\beta}_{nm,ij}$
as the matrix elements of $Q$. The NL$\sigma$M action is then invariant
under unitary transformations. We are interested in symmetry breaking
operators that are quadratic in $Q$. This case was first considered by
Pruisken.\cite{Pruisken} There are two irreducible representations that
correspond to symmetrized and antisymmetrized products of the $Q$. Any
operator
\bml
\label{eqs:B.1}
\be
O = \int d{\bf x}\sum_{1234} v_{12,34}\,Q_{12}({\bf x})\,Q_{34}({\bf x})\quad,
\label{eq:B.1a}
\ee
should thus be written as
\be
O = O_{+} + O_{-}\quad,
\label{eq:B.1b}
\ee
with
\bea
O_{\pm} &=& \frac{1}{2}\int d{\bf x}\sum_{12,34} v_{12,34}\,\left[
          Q_{12}({\bf x})\,Q_{34}({\bf x})\right.
\nonumber\\
&&\hskip 90pt\left. \pm Q_{32}({\bf x})\,Q_{14}({\bf x})
          \right]\quad.
\label{eq:B.1c}
\eea
\eml%
Here $1\equiv (n_1,\alpha_1,i_1)$, etc.
$O_{+}$ and $O_{-}$ require one renormalization constant each, so two
additional renormalization constants are needed to renormalize the
NL$\sigma$M with arbitrary massive insertions of order $Q^2$.

A complication lies in the fact that in the present model, the coupling
constants $H^{(1)}$ and $H^{(2)}$ multiply frequency dependent terms,
and the frequency gets integrated over in perturbation theory. As a
result, ratios of the $J$ and $H$ appear in perturbation theory, and
the proof given in Refs.\ \onlinecite{soft_insertions} does not apply.
This is true a fortiori in the case of an instantaneous electron-electron
interaction, where the additional operators are not even massive
insertions. Nevertheless, while no actual proof of renormalizability 
exists in this case, Ref.\ \onlinecite{us_NPB} has presented 
substantial evidence from perturbation
theory that the model is still renormalizable with two additional
renormalization constants for the interaction. The same conclusion is
expected to hold in the annealed disorder case. 

\section{Comparison with the case of an instantaneous interaction}
\label{app:C}

In this appendix we compare the flow equations derived in 
Sec.\ \ref{subsec:IV.B} with those for the case of an instantaneous
electron-electron interaction.

In the instantaneous interaction case one has spin-singlet and spin-triplet
interactions amplitudes $K_{\rm s}$ and $K_{\rm t}$, that
are analogous to $J^{\rm (1,s)}$ and $J^{\rm (1,t)}$, respectively.
The analog of $J^{\rm (2,t)}$ does not exist. Instead of the two
frequency couplings $H^{(1)}$ and $H^{(2)}$ there is only one coupling
constant $H$, which is proportional to the specific heat coefficient.
$\partial n/\partial\mu$ and $\chi_{\rm s}$ are proportional to
$H + K_{\rm s}$ and $H + K_{\rm t}$, respectively.\cite{us_R}

In the absence of magnetic impurities, a magnetic field, or spin-orbit
scattering, $K_{\rm t}$ flows towards large values, and after some
transient behavior the one-loop flow equations take the form
\bml
\label{eqs:C.1}
\bea
\frac{dg}{d\ell} &=& -\epsilon g + O(g^3)\quad,
\label{eq:C.1a}\\
\frac{dh}{d\ell} &=& \frac{3}{8}\,gk_{\rm t} + O(g^2)\quad,
\label{eq:C.1b}\\
\frac{dk_{\rm s}}{d\ell} &=& \frac{-3}{8}\,gk_{\rm t} + O(g^2)\quad,
\label{eq:C.1c}\\
\frac{dk_{\rm t}}{d\ell} &=& \frac{1}{2}\,gk_{\rm t}^2/h\quad.
\label{eq:C.1d}
\eea
\eml%
A comparison with Eqs.\ (\ref{eqs:4.13}) shows that the two behaviors
are very similar, except that in the instantaneous interaction case
$k_{\rm t}$ flows to infinity much faster than $-k_{\rm s}$. In
particular, the conductivity and $\partial n/\partial\mu$ are not
renormalized in either case (and neither is the density of states),
while the magnetic susceptibility and the specific heat coefficient
both diverge, albeit the latter only logarithmically in the
instantaneous interaction case.\cite{us_IFS} Strictly at one-loop order, 
the physical interpretation of the RG flow was long considered not obvious, 
as has been stressed in the literature many times. However, the analysis 
given in Ref.\ \onlinecite{us_IFS}, combined with the
detailed discussion of the ferromagnetic transition in 
Ref.\ \onlinecite{us_fm_local}, has shown that the proper interpretation
is in terms of a ferromagnetic transition in $d=2+\epsilon$, as has been 
discussed in Sec.\ \ref{subsec:V.A}.

\section{Perturbation theory for the free energy}
\label{app:D}

Here we calculate the free energy in perturbation theory. This serves
as a check on our scaling arguments for various observables in 
Sec.\ \ref{subsec:V.A}.

To zeroth order in a loop expansion, the free energy density $f$ is given by
the saddle-point action. This yields free-electron values for all
thermodynamic quantities. The first correction, $\Delta f$, is obtained by 
integrating over the fields $q$ in Gaussian approximation. From
Eqs.\ (\ref{eqs:4.0}) we find
\bml
\label{eqs:D.1}
\be
\Delta f = \Delta f_{\rm s} + 3\Delta f_{\rm t}\quad.
\label{eq:D.1a}
\ee
Here
\bea
\Delta f_{\rm s,t} &=& \frac{iG}{H^{(1)}}\,J_{\rm s,t}\int_0^1 d\eta\,
   (H^{(2)} + \eta J_{\rm s,t})\,
\nonumber\\
&&\times \frac{1}{V}\sum_{\rm k}\int_0^{\infty}
   d\omega\ n(\omega/T)\,{\cal D}^{\rm s,t}({\bf k},\omega,T)\quad,
\label{eq:D.1b}
\eea
with (cf. Eq.\ (\ref{eq:4.1e}))
\be
{\cal D}^{\rm s,t}({\bf k},\omega,T) = \frac{1}{{\bf k}^2 - iGH^{(1)}\omega
   + G(H^{(2)} + \eta J_{\rm s,t})2\pi T}\quad.
\label{eq:D.1c}
\ee
The function
\be
n(x) = \frac{1}{2}\,\coth (\frac{x}{2}) - \frac{1}{x}\quad,
\label{eq:D.1d}
\ee
\eml%
serves as a convenient means for transforming the sum over Matsubara
frequencies into a real-frequency integral, and we have used the familiar
``charging formula'' trick of integrating over the interaction constants
in order to improve convergence.


\begin{references}
\b{us_R} See, e.g., D. Belitz and T.R. Kirkpatrick, Rev. Mod. Phys. {\bf 66},
 261 (1994).
\b{LM_footnote} This is the local moment concept that was considered in a
 classical context by Viktor Dotsenko, A.B. Harris, D. Sherrington, and
 R.B. Stinchcombe, J. Phys. A {\bf 28}, 3093 (1995), and that was generalized
 to quantum field theories in Refs.\ \onlinecite{us_rr_magnets} and 
 \onlinecite{us_letter}. Other, more
 spatially localized, local moment concepts have also frequently been
 discussed in the literature, see R.N. Bhatt and P.A. Lee, Phys. Rev. Lett.
 {\bf 48}, 344 (1982); R.N. Bhatt and D.S. Fisher, Phys. Rev. Lett. {\bf 68},
 3072 (1992). In the context of the disordered interacting electron problem
 that we will frequently refer to, local moments have been discussed by
 C. Castellani, C. DiCastro, P.A. Lee, M. Ma, S. Sorella, and E. Tabet,
 Phys. Rev. B {\bf 30}, 1596 (1984); A.M. Finkelstein, Pisma Zh. Eksp. Teor.
 Fiz. {\bf 40}, 63 (1984) [JETP Lett. {\bf 40}, 796 (1984); and more recently
 by A.V. Andreev and A. Kamenev, Phys. Rev. Lett. {\bf 81}, 3199 (1998); and
 by B. N. Narozhny, I. L. Aleiner, and A. I. Larkin, Phys. Rev. B {\bf 62},
 14898 (2000).
\b{us_rr_magnets} R. Narayanan, T. Vojta, D. Belitz, and T.R. Kirkpatrick,
 Phys. Rev. B {\bf 60}, 10150 (1999).
\b{us_letter} D. Belitz, T.R. Kirkpatrick, and Thomas Vojta, Phys. Rev. Lett.
 {\bf 84}, 5176 (2000).
\b{Mott} For a general discussion and classification of metal-insulator
 transitions, see, N.F. Mott, {\it Metal-Insulator Transitions}, 
 Taylor \& Francis (London 1990).
\b{us_local_OP} D. Belitz, T.R. Kirkpatrick, and Thomas Vojta, 
 cond-mat/0109547 (Phys. Rev. B, in press).
\b{us_chi_s} D. Belitz, T.R. Kirkpatrick, and Thomas Vojta, Phys. Rev. B
 {\bf 55}, 9452 (1997).
\b{NLsM_footnote} This is true within the NL$\sigma$M description of the
 MIT, where the Goldstone modes in the ordered (i.e., metallic) phase are
 also the soft modes that drive the transition. This results in the somewhat
 surprising conclusion that, for any given observable, there is a relation
 between its critical behavior
 at the critical fixed point and its corrections to scaling at the stable
 fixed point that describes the ordered phase. This is true for any
 NL$\sigma$M, including the one for the classical Heisenberg 
 ferromagnet.\cite{ZJ} Related considerations have at times led to a debate
 as to whether the ferromagnetic transition found within the NL$\sigma$M in 
 $d=2+\epsilon$ and the one found within $\phi^4$-theory in $d=4-\epsilon$
 are really physically identical; see E. Br{\'e}zin and S. Hikami,
 cond-mat/9612016, and references therein.
\b{ZJ} See, J. Zinn-Justin, {\it Quantum Field Theory and Critical Phenomena},
 Clarendon Press (Oxford 1989), and references therein.
\b{us_IFS} T.R. Kirkpatrick and D. Belitz, J. Phys. Cond. Matt. {\bf 2}, 5259
 (1990); Phys. Rev. B {\bf 45}, 3187 (1992). In these papers, the precise
 nature of the first phase transition was unclear. The identification as a
 ferromagnetic transition was provided by T.R. Kirkpatrick and D. Belitz,
 Phys. Rev. B {\bf 53}, 14364 (1996), and in Ref.\ \onlinecite{us_fm_local}.
\b{ItzyksonDrouffe} See, e.g., C. Itzykson and J.-M. Drouffe, {\it Statistical
 Field Theory}, Cambridge University Press (Cambridge 1989).
\b{Grinstein} See, e.g., G. Grinstein, in {\it Fundamental Problems in
 Statistical Mechanics VI}, E.G.D. Cohen (ed.), North Holland (Amsterdam,
 1985).
\b{F} A.M. Finkel'stein, Zh. Eksp. Teor. Fiz. {\bf 84}, 168 (1983) [Sov.
 Phys. JETP {\bf 57}, 97 (1983)].
\b{Wegner} F. Wegner, Z. Phys. B {\bf 35}, 207 (1979). We use the fermionic
 formulation of the model given in Ref.\ \onlinecite{ELK}. See also
 Ref.\ \onlinecite{us_fermions}.
\b{ELK} K.B. Efetov, A.I. Larkin, and D.E. Khmelnitskii, Zh. Eksp. Teor. Fiz.
 {\bf 79}, 1120 (1980) [Sov. Phys. JETP {\bf 52}, 568 (1980)].
\b{us_fermions} D. Belitz and T.R. Kirkpatrick, Phys. Rev. B {\bf 56}, 6513
 (1997); D. Belitz, T.R. Kirkpatrick, and F. Evers, Phys. Rev. B {\bf 58},
 9710 (1998).
\b{WilsonKogut} K.G. Wilson and J. Kogut, Phys. Rep. {\bf 12}, 75 (1974).
\b{Pruisken} A.M.M. Pruisken, Phys. Rev. B {\bf 31}, 416 (1985).
\b{us_NPB} D. Belitz and T.R. Kirkpatrick, Nucl. Phys. B {\bf 316}, 509 (1989).
\b{us_fm_local} D. Belitz, T.R. Kirkpatrick, Maria Teresa Mercaldo, and Sharon
 L. Sessions, Phys. Rev. B {\bf 63}, 174427 (2001); {\it ibid.} B {\bf 63},
 174428 (2001).
\b{BrinkmanRice} W.F. Brinkman and T.M. Rice, Phys. Rev. B {\bf 2}, 4302 
 (1970).
\b{TV_RN} Thomas Vojta and R. Narayanan, private communication.
\b{F2} A.M. Finkel'stein, Zh. Eksp. Teor. Fiz. {\bf 86}, 367 (1984) [Sov.
 Phys. JETP {\bf 59}, 212 (1984)].
\b{Weinberg} S. Weinberg, {\it The Quantum Theory of Fields}, Vol. 1, Cambridge
 University Press (Cambridge 1995), Sec. 12.4, and references therein.
\b{BrezinZJ} E. Br{\'e}zin, J. Zinn-Justin, and J.C. Le Guillou, Phys. Rev. D
 {\bf 14}, 2615 (1976).
\b{soft_insertions} E. Br{\'e}zin, J. Zinn-Justin, and J.C. Le Guillou, Phys. 
 Rev. B {\bf 14}, 4976 (1976); D. H{\"o}f and F. Wegner, Nucl. Phys. B {\bf 275}
 [FS17], 561 (1986); F. Wegner, Nucl. Phys. B {\bf 280} [FS18], 193 (1987);
 {\it ibid.} {\bf 280} [FS18], 210 (1987). The first of these papers  uses the
 field theory terminology ``soft insertions'' for the terms in question.
\end{references}
\end{document}